\documentclass[journal,10pt]{IEEEtran}
\usepackage{cite}
\usepackage{etoolbox}
\usepackage{mathtools, cuted}
\usepackage{tabu}
\usepackage{array}
\usepackage{comment}

\ifCLASSINFOpdf
\else
\fi

\usepackage{amsmath, amsthm, amssymb, amsfonts}

\usepackage{amsmath}
\usepackage{graphicx}
\usepackage{epstopdf}
\usepackage{color}

\makeatletter
\DeclareRobustCommand\bfseriesitshape{%
\not@math@alphabet\itshapebfseries\relax
\fontseries\bfdefault
\fontshape\itdefault
\selectfont
}
\makeatother

\DeclareTextFontCommand{\textbfit}{\bfseriesitshape}

\begin{document}

%
\title{On the Secrecy Performance of Generalized User Selection for Interference-Limited Multiuser Wireless Networks}

\author{\IEEEauthorblockN{Yazan H. Al-Badarneh, Costas N. Georghiades, Redha M. Radaydeh, Mohamed-Slim Alouini} 
%
%
%
%
%
%

}

\maketitle 



\begin{abstract}

We investigate the secrecy performance of a multiuser diversity scheme for an interference-limited wireless network with a base-station (BS), $N$ legitimate users and an eavesdropper, in the presence of a single dominant interferer. Assuming interference dominates noise power at the eavesdropper and at each legitimate user's receiver, the BS transmits information to the legitimate user with the $k$-th best (highest) signal-to-interference ratio. We derive a closed-form expression for the secrecy outage probability for an arbitrary $N$ and an asymptotic expression for a fixed $k$ and large $N$. Furthermore, we derive a closed form asymptotic expression for the ergodic secrecy capacity of the $k$-th best user and show that it scales like $O\left(\log(N)\right)$ for a fixed $k$ and large $N$. 


\end{abstract}
\section{Introduction} 
The notion of secure communication was first introduced by Shannon in \cite{6769090}. Thereafter, Wyner introduced  the wiretap channel in which Alice transmits confidential messages to Bob in the presence of an eavesdropper, Eve \cite{6772207}. Physical layer (PHY) security was first investigated in \cite{4529264}, where the authors analyze the secrecy outage probability (SOP) and  the ergodic secrecy capacity (ESC) for single-input single-output (SISO) systems subject to a quasi-static Rayleigh fading. 

Multiuser diversity can improve PHY security \cite{7562282} \cite{7517396}. The impact of interference on the PHY security for multiuser diversity schemes is investigated in \cite{7259422} where the authors analyze the SOP and secrecy diversity order of multiuser diversity scheduling in the presence of cochannel interference, but no closed form expression was derived for the SOP. In \cite{6985747}, the SOP of a multiuser diversity scheme of cognitive radio systems is investigated in the presence of interference from the primary transmitter. The fading statistics of the interference were modeled as complex Gaussian, assuming the primary signal is generated by a random Gaussian codebook. Most recently, the effect of fading of multiple interference channels is considered in \cite{8016351} where the authors analyze the SOP for a single user (no multiuser diversity).

The mentioned previous works have only focused on the secrecy performance of the conventional multiuser diversity scheme where the user with the best fading (the best user) is selected.  However, in a practical wireless network the best user may not be selected under given traffic conditions. This might happen when the best user is unavailable or occupied by other service requirements \cite{LI2015745}, in handoff situations \cite{6146496} or due to scheduling delay \cite{6171803}. Accordingly, the main contribution of this paper is to study the secrecy performance of the $k$-th best user selection scheme, which is a generalized selection criterion that includes the best user (i.e., $k=1$) as a special case. In particular, we analyze the SOP and ESC of a $k$-th best user selection scheme of a multiuser wireless network in the presence of a single dominant interferer \footnote{The single dominant interferer assumption has been widely considered in the literature, see \cite{1532219}, \cite{6365852} and references therein. The interfering BS here can be viewed as a co-channel interferer serving another cell (network) that causes an interference to the cell  of interest.}.  Assuming that the noise power at each user's receiver and Eve's receiver are negligible compared to the interference power,  and the user with the $k$-th best signal to interference ratio (SIR) is selected from a total number of users $N$, we derive a closed-form expression for the secrecy outage probability for an arbitrary $N$ and an asymptotic expression for a fixed $k$ and large $N$. Furthermore, we derive an asymptotic closed form expression for the ESC of the $k$-th best user and show that the ESC scales like $O\left(\log(N)\right)$ for a fixed $k$ and large $N$. 



In Section II we discuss the system model. In Section III we analyze the SOP of the $k$-th best user and in Section IV the ESC. Sections V and VI present numerical results and the conclusion, respectively.

\section{System model}
As shown in Fig. 1, we consider a wireless network consisting of one BS (Alice), $N$ legitimate users (Bobs) and an eavesdropper (Eve), in the presence of another interfering BS.  At any time instant, only one legitimate user is scheduled to receive confidential messages from Alice.  We assume that Eve is equipped with $L$ receive antennas and all other terminals with one antenna each. Let $h_{i}$ and $t_{l}$ denote the channel gain from Alice to the $i$-th user's receiver and Eve's $l$-th receive antenna, respectively. Let $g_{i}$ denote the channel gain from the interfering BS to the $i$-th user's receiver and $e_{l}$ the channel gain from the interfering BS to Eve's $l$-th receive antenna. The channel gains are modeled as independent Rayleigh distributed random variables. In particular, $|t_{l}|^{2}$ and $|e_{l}|^{2}$, for $l=1, 2,..., L$, are independent identically distributed (i.i.d) exponential random variables with parameters $\lambda_{E}$ and $\beta_{E}$, respectively. Furthermore, $|h_{i}|^{2}$ and $|g_{i}|^{2}$, for $i=1, 2,..., N$, are i.i.d exponential random variables with parameters $\lambda_{M}$ and $\beta_{M}$, respectively. 
\begin{figure}[h]\label{fig:1}
\begin{center}
\includegraphics[width=1\columnwidth]{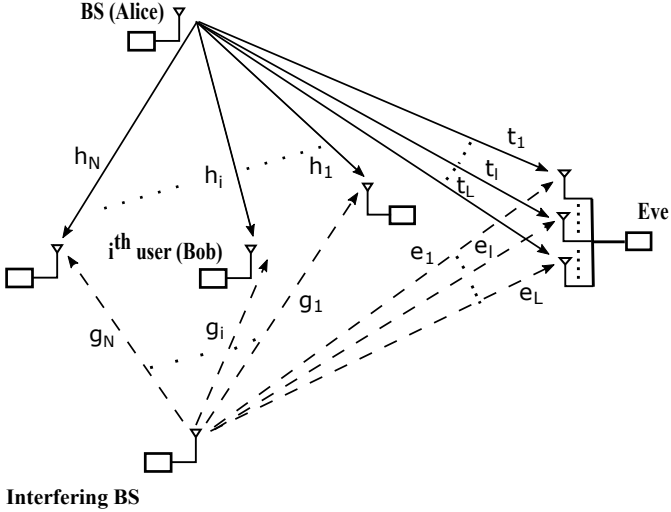}
\caption{ Multiuser wireless network with a BS (Alice), $N$ legitimate users (Bobs), an eavesdropper (Eve) equipped with $L$ antennas, in the presence of another interfering BS.}
\end{center}
\end{figure} 
Assuming the interference power from the BS is much larger than the noise power at the $i$-th user's receiver, the signal-to-interference ratio (SIR) at the $i$-th user's receiver is given by 
\begin{equation} \label{eq:1}
Z_{i}=\frac{P |h_{i}|^{2}}{P_{I} |g_{i}|^{2}}, 
\end{equation}
where $P$ and $P_{I}$ are the transmit power of Alice and the interfering BS, respectively. The cumulative distribution function (CDF)  $Z_{i}$ is given by \cite{6924725} 
\begin{equation}\label{eq:2}
F(z)= \frac{z}{C_{M}+ z} \ u(z), 
\end{equation}
where $C_{M}=\frac{P \beta_{M} }{P_{I}\lambda_{M} }$ and $u(z)$ is the unit step function. We sort the random variables $Z_{i}$ in an increasing order denoted as $Z_{(1)} \leq Z_{(2)}.... \leq Z_{(N-k+1)} \leq .... \leq Z_{(N)}$,  such that Alice selects the user with the $k$-th highest SIR, $Z_{(N-k+1)}$. The CDF of $Z_{\left(N-k+1\right)}$ then is \cite{david2003order}
\begin{equation}\label{eq:3}
F_{Z_{\left(N-k+1\right)}}(x)=\sum_{v=N-k+1}^{N} \binom{N}{v} (F(z))^{v} \left(1-F(z) \right)^{N-v}. 
\end{equation}

Assuming Eve is equipped with $L$ receive antennas and the noise power at the $l$-th receive antenna is negligible compared to the interference power from the interfering BS, the instantaneous SIR at the Eve's $l$-th receive antenna is $X_{l} =\frac{P |t_{l}|^{2}}{P_{I} |e_{l}|^{2}}$. The CDF of $X_{l}$ is similar to that in (\ref{eq:2}), but with parameter $C_{E}=\frac{P \beta_{E}}{ P_{I}\lambda_{E}}$. Assuming that selection combining (SC) is employed at Eve, such that the best receive antenna is selected, the instantaneous SIR of the SC output is $ X_{(L)} =\underset{k=1, ..., L }{\text{max}}X_{l}$. Using (\ref{eq:3}), the CDF and the probability density function (PDF) of $X_{(L)}$ are 
\begin{equation}\label{eq:4}	
F_{X_{(L)}}(z)= \left( \frac{z}{C_{E}+ z} \right)^{L} \ u(z), 
\end{equation}
\begin{equation}\label{eq:5}
f_{X_{(L)}}(z)= \frac{d \left(F_{X_{(L)}}(z)\right)}{dz}= \frac{L C_{E} \ z^{L-1}}{(C_{E}+ z)^{L+1}} \ u(z). 
\end{equation}

\section{ Secrecy Outage Probability (SOP)} 
In this section we focus on analyzing the secrecy outage probability assuming that Alice has no knowledge about the eavesdropper's channel state information (CSI), similar to  \cite{4529264}, \cite{7556417}.  

For $k$-th best user selection, the secrecy capacity is given by \cite{4529264}
\begin{gather} \label{eq:6}
\begin{split}
 C_{s}(k,N,L)=
\begin{cases}
\log_{2}\left(\frac{1+Z_{\left(N-k+1\right)}}{1+ X_{(L)}} \right), \ Z_{\left(N-k+1\right)}>X_{(L)} \\ 
0, \ \ \ \ \ \ \ \ \ \ \ \ \ \ \ \ \ \ \ \ \ \ \ Z_{\left(N-k+1\right)} \leq X_{(L)}\\
\end{cases} .
\end{split}
\end{gather}
The SOP for a target secrecy rate $R_{s}$ is given by \cite{4529264}
\begin{gather}\label{eq:7}
\begin{split}
P_{out}(R_{s})&= {\rm Pr} \left\{  C_{s}(k,N,L)\leq R_{s} \right\}\\
&=\int_{0}^{\infty} f_{X_{(L)}}(z) F_{Z_{\left(N-k+1\right)}}\left (\tau-1+ \tau z\right) dz,
\end{split}
\end{gather}
where $ \tau = 2^{R_{s}}$. Using (\ref{eq:7}), the probability of strictly positive secrecy capacity (SPSC) can be evaluated as ${\rm Pr} \left\{C_{s}(k,N,L) > 0 \right\}=1-P_{out}(0)$.

We derive next an exact expression for the SOP of $k$-th best user selection for arbitrary values of $N$ and $L$, an asymptotic expression for the SOP of the $k$-th best user for arbitrary $L$ and large $N$ compared to fixed $k$ and $\tau$, and obtain a simple asymptotic expression for the SOP for large $N$ compared to fixed $k$ and $\tau$ and for large $L$. \\   


\noindent{\textit{Proposition 1:}} For arbitrary $N$ and $L$, the exact SOP of the $k$-th best user is
\begin{gather}\label{eq:8}
\begin{split}
P_{out}(R_{s}) =& \frac{L}{ {\tau^{L}\left(C_{E}\right)}^{L}} \sum_{v=N-k+1}^{N} \binom{N}{v} \ {\left(C_{M}\right)}^{N-v} \sum_{j=0}^{v} \binom{v}{j}  \\ 
& \times \left( \tau-1\right)^{v-j} B \left( L+j,N-j+1 \right) \\
& \times  { \left( {\it \tau-1} +C_{M} \right) ^{L+j-N} } {}_2F_1\left( L+1,L+j;  \right.\\ 
&\left. N+L+1; 1-{\frac {{\it \tau-1 }+C_{M}}{\tau\,C_{E}}} \right),
\end{split}
\end{gather}
where $\tau = 2^{R_{s}}$, ${}_2F_1\left( x,y; z;w\right)$ is the Gauss hypergeometric function and $B \left( x,y \right)$ is the Beta function. 

\noindent \textit{Proof:} Using (\ref{eq:2}), (\ref{eq:3}) and (\ref{eq:5}), after some basic algebraic
manipulations, $P_{out}(R_{s})$ in (\ref{eq:7}) can be expressed as 
\begin{gather}\label{eq:9}
\begin{split}
P_{out}(R_{s}) =& L C_{E} \sum_{v=N-k+1}^{N} \binom{N}{v} \tau^{v- N} \ {\left(C_{M}\right)}^{N-v} \\
& \times \underbrace{ \int_{0}^{\infty} {\frac { z^{L-1} \left( z+\it{\frac{\tau-1}{\tau} }\right) ^{v} }{\left( z+ \frac{ {\it \tau-1} +C_{M}}{\tau} \right) ^{N} \left( z+C_{E} \right) ^{L+1}}} dz}_{I}. 
\end{split}
\end{gather}
Applying binomial expansion for the term $ \left( z+\it{\frac{\tau-1}{\tau}}\right) ^{v}$ and making use of Eq. (3.197.1) of \cite{jeffrey2007table}, ${I}$ can be expressed as 
\begin{gather}\label{eq:10}
\begin{split}
I=&\sum_{j=0}^{v} \binom{v}{j} \left( \frac{\tau-1}{\tau}\right)^{v-j} {\frac {B \left( L+j,N-j+1 \right) }{{C_{E}}^{L+1}} }   \\
& \times \left( \frac{ {\it \tau-1} +C_{M}}{\tau} \right) ^{L+j-N} {}_2F_1\left( L+1,L+j;   \right.\\ 
&\left. N+L+1; 1-{\frac {{\it \tau-1 }+C_{M}}{\tau\,C_{E}}} \right). 
\end{split}
\end{gather}
Combining (\ref{eq:9}) and (\ref{eq:10}), $P_{out}(R_{s})$ can be finally expressed as in (\ref{eq:8}). Setting $R_{s}=0$ (i.e., $\tau=1$) in (\ref{eq:8}), the probability of SPSC is 
\begin{gather}\label{eq:11}
\begin{split}
{\rm Pr} \left\{ C_{s}(k,N,L) > 0 \right\}=& 1- L \left( \frac{C_{M}}{C_{E}} \right)^L \sum_{v=N-k+1}^{N} \binom{N}{v}\\
 & \times B \left( L+v,N-v+1 \right) { } {}_2F_1\left( L+1,  \right.\\ 
&\left. L+v;  N+L+1; 1-{\frac {C_{M}}{C_{E}}} \right). 
\end{split}
\end{gather}

\noindent{\textit{Proposition 2:}} For arbitrary $L$ and large $N$, with respect to fixed $k$ and $\tau$, the SOP of the $k$-th best user can be approximated as
\begin{gather}\label{eq:12}
\begin{split}
P_{out}(R_{s})\approx 1- \left( \frac {b_{N}}{\tau\,C_{E}}\right)^{k} U\left(k; k+1-L;{\frac {b_{N}}{\tau\,C_{E}} }\right), 
\end{split}
\end{gather}
where $b_{N}= C_{M} (N-1)$ and $\textit{U}\left( a;d;z \right)$ is the Tricomi hypergeometric function \cite{1576535}. 



\noindent{\textit{Proof:}} As shown in Proposition 2 of \cite{yazanVTC2018}, if the random variable $Z_{i}$ has a CDF $F(z)$ as in (\ref{eq:2}), then for a fixed $k$ and $N \to \infty$, $\frac{Z_{(N-k+1)}}{b_{N}}$ converges in distribution to a random variable $Z$ whose CDF, $G^{k}(z)$, has an inverse gamma distribution. 
\begin{eqnarray}\label{eq:13}
G^{k}(z)= \frac{\Gamma \left( k,{\frac {1}{z}} \right)}{(k-1)!} u(z).  
\end{eqnarray}
$\Gamma(s,x)= \int_{x}^{\infty} t^{s-1} e^{-t} dt$ is the upper incomplete gamma function and $ b_{N}= C_{M} (N-1)$. Equivalently, for fixed $k$ and $ N \to \infty$, $F_{Z_{(N-k+1)}}(z)$ can be approximated as
\begin{eqnarray}\label{eq:14} 
F_{Z_{(N-k+1)}}(z)\approx \frac{\Gamma \left( k,{\frac {b_{N}}{z}} \right)}{(k-1)!} u(z).
\end{eqnarray}

Based on the asymptotic distribution of $ \frac{Z_{(N-k+1)}}{b_{N}}$ above  and noting that $b_{N}$ is an increasing function of $N$, we derive next an asymptotic expression for $P_{out}(R_{s})$. Invoking (\ref{eq:7}), we have
\begin{gather} \label{eq:15}
\begin{split}
P_{out}(R_{s})&=\int_{0}^{\infty} f_{X_{(L)}}(z) F_{Z_{\left(N-k+1\right)}}\left (\tau-1+ \tau z\right) dz\\
&={\rm Pr} \left\{ Z_{\left(N-k+1\right)} \leq \tau-1+ \tau X_{(L)}  \right\}\\
&={\rm Pr}\left\{ \frac{Z_{\left(N-k+1\right)}}{b_{N}} \leq \frac{\tau-1+ \tau X_{(L)}}{b_{N}}  \right\}\\ 
& \approx {\rm Pr}\left\{Z \leq \frac{\tau X_{(L)}}{b_{N}} \right\},\\ 
\end{split}
\end{gather}
 for  fixed $k$ and $\tau$ and $ N \to \infty$, where the CDF of $Z$ is as in (\ref{eq:13}).  Using (\ref{eq:15}), $P_{out}(R_{s})$ can be expressed as 
\begin{gather}\label{eq:16}
\begin{split}
P_{out}(R_{s})\approx \underbrace{\int_{0}^{\infty} \frac{L C_{E} z^{L-1}}{(C_{E}+ z)^{L+1}} \frac{ \ \Gamma \left( k,{\frac {b_{N}}{\tau z}} \right) }{(k-1)!} dz}_\text{$I_{1}$}, 
\end{split}
\end{gather}
as $ N \to \infty$ and for fixed $k$ and $\tau$. Using integration by parts:
\begin{gather}\label{eq:17}
\begin{split}
I_{1}&= 1- \int_{0}^{\infty} \left( {\frac {z}{z+C_{E}}} \right) ^{L} {\frac {{(b_{N})}^{k}{\tau}^{-k}{z}^{-1-k}}{ \left( k-1 \right) !}{{ e}^{-{\frac {b_{N}}{\tau\,z}}}}} dz.
\end{split}
\end{gather}
Using $u= \frac{C_{E}}{z} $ and Eq. (39) of \cite{1576535}, $I_{1}$ can be finally expressed as in (\ref{eq:12}). Then, the probability of SPSC is  
\begin{gather}\label{eq:18}
\begin{split}
{\rm Pr} \left\{ C_{s}(k,N,L) > 0 \right\} \approx \left( \frac {b_{N}}{\,C_{E}}\right)^{k} U\left(k; k+1-L;{\frac {b_{N}}{\,C_{E}} }\right). 
\end{split}
\end{gather} 
In Proposition 2 above, we derive the SOP of the $k$-th best user for an arbitrary $L$ and large $N$ relative to fixed $k$ and $\tau$. If we further assume that $L$ is large we derive a simpler and an accurate expression for the SOP as in Proposition 3 below. \\


\noindent{\textit{Proposition 3:}} For large $N$ compared to fixed $k$ and $\tau$ and for large $L$, the SOP of the $k$-th best user can be approximated as
\begin{gather}\label{eq:19}
\begin{split}
P_{out}(R_{s})\approx 1- \left(1+ \frac{\tau b_{L}}{b_{N}} \right)^{-k},
\end{split}
\end{gather}
where  $ b_{N}= C_{M} (N-1)$ and $ b_{L}= C_{E} (L-1)$. 
 
\noindent{\textit{Proof:}} As discussed earlier, $Z_{i}$ and $Z_{l}$ have the same CDF  with parameters  $C_{M}$ and $C_{E}$, respectively. Then the asymptotic distribution of $X_{(L)}$ for large $L$ can be obtained by replacing $N$ with $L$ and setting $k=1$ in (\ref{eq:14}). Hence, as $L \to \infty $, with $b_{L}=C_{E} (L-1)$, $F_{X_{(L)}}(z)$ can be expressed as 
\begin{eqnarray}\label{eq:20}
F_{X_{(L)}}(z)\approx  e^{-\frac {b_{L}}{z}} u(z).
\end{eqnarray}
Making use of (\ref{eq:20}) in (\ref{eq:15}) we have 
\begin{gather} \label{eq:21}
\begin{split}
P_{out}(R_{s})& \approx {\rm Pr}\left\{Z \leq \frac{\tau X_{(L)}}{b_{N}} \right\}\\
&= \int_{0}^{\infty}  \frac{d \left( e^{-\frac {b_{L}}{z}}  \right)}{dz}   \frac{ \ \Gamma \left( k,{\frac {b_{N}}{\tau z}} \right) }{(k-1)!} dz\\
&=\int_{0}^{\infty}  \frac{b_{L} \ e^{-\frac {b_{L}}{z}}}{z^{2}}   \frac{ \ \Gamma \left( k,{\frac {b_{N}}{\tau z}} \right) }{(k-1)!} dz\\
&= 1- \left(1+ \frac{\tau b_{L}}{b_{N}} \right)^{-k},
\end{split}
\end{gather}
for large $N$ relative to fixed $k$ and $\tau$ and for large $L$. The above integral is evaluated using Eq. (6.451, 2) of \cite{jeffrey2007table}. $P_{out}(R_{s})$  in (\ref{eq:21}) is an increasing function of $L$ and $k$, and a decreasing function of $N$. As a special case, if $N=L$ and  $L$ is large, $P_{out}(R_{s})$ converges to a constant value, i.e,. $P_{out}(R_{s})\approx 1- \left(1+ \frac{\tau \beta_{E} \lambda_{M}}{\lambda_{E} \beta_{M}} \right)^{-k}$. This shows that if $L$ is large and scales linearly with $N$, the SOP converges to a constant that only depends on the fading parameters and $R_{s}$. This can be intuitively explained by the fact that for large $L$ and scaling linearly with $N$, the multiuser diversity effect on the SOP is eliminated due to the  employment of the selection combining scheme at Eve. 

\section{ Ergodic secrecy capacity (ESC) } 
In the case where the eavesdropper's CSI is available at Alice (active eavesdropping), the ESC is essentially a fundamental secrecy performance metric \cite{4529264}, \cite{6908024}.  

If the eavesdropper's CSI is available at Alice, the ESC  for $k$-th best user selection can be expressed as \cite{6908024}
\begin{gather}\label{eq:22}
\begin{split}
\overline{C_{s}}{(k,N,L)} &=\frac{1}{\ln(2)}\int_{0}^{\infty} \frac{F_{X_{(L)}}(z)}{1+z} \left( 1- F_{Z_{\left(N-k+1\right)}}\left(z \right) \right) dz,
\end{split}
\end{gather}
which in general is intractable to express in closed form for arbitrary values of $N$ and $L$. However, using the asymptotic approximation of $F_{Z_{(N-k+1)}}(z)$ in (\ref{eq:14}) yields a closed form asymptotic expression for the ESC when Eve has a single antenna (i.e., $L=1$). In Proposition 4 below, we derive the ESC of the $k$-th best user for large $N$ relative to a fixed $k$ and $L=1$. We use $\overline{C_{s}}{(k,N)}$ to denote $\overline{C_{s}}{(k,N,L)}$ at  $L=1$.   \\

\noindent{\textit{Proposition 4:}} For large $N$ relative to fixed $k$, the ESC of the $k$-th best user can be approximated as
\begin{gather} \label{eq:23}
\begin{split}
\overline{C_{s}}{(k,N)} \approx
\begin{cases}
\frac{-\psi(k) +{\frac{C_{E} V\left(k;\frac { b_{N}}{C_{E}}\right)-V\left(k;b_{N}\right)}{C_{E}-1}  }}{\ln(2)},  \ \ \ \ \ \ \ \ \ \ C_{E} \neq 1\\
\frac{-\psi(k)+ V\left(k;b_{N}\right) -{\left(b_{N}\right)}^{k} {{ e}^{b_{N}}}\Gamma  \left( -k+1,b_{N} \right)}{\ln(2)},   C_{E} = 1
\end{cases}, 
\end{split}
\end{gather}
where $\psi(k)$ is the digamma function,  $b_{N}= C_{M} (N-1)$ and  $V(k;a)$ is as expressed in (\ref{eq:30}).  

\noindent{\textit{Proof:}} Using (\ref{eq:14}) and the fact that $ \frac{\Gamma \left( k,x\right)}{(k-1)!}=1-\frac{\gamma \left( k,x\right)}{(k-1)!}$, then 
\begin{eqnarray}\label{eq:24}
1-F_{Z_{(N-k+1)}}(z)\approx \frac{\gamma \left( k,{\frac {b_{N}}{z}} \right)}{(k-1)!}, 
\end{eqnarray}
where $\gamma(s,x)= \int_{0}^{x} t^{s-1} e^{-t} dt$ is the lower incomplete gamma function. Substituting (\ref{eq:4}) and (\ref{eq:24}) in (\ref{eq:22}), yields 
\begin{gather}\label{eq:25}
\begin{split}
\overline{C_{s}}{(k,N)}  \approx \frac{1}{\ln(2)} \underbrace{ \int_{0}^{\infty} \frac{z}{(1+z)(C_{E}+z)} \frac{ \gamma \left( k,{\frac {b_{N}}{z}} \right)}{ (k-1)!} dz}_\text{$I_{2}$}.  
\end{split}
\end{gather}
Using $x= \frac{b_{N}}{z} $ and the integral representation of the lower incomplete gamma function, $I_{2}$ can be expressed as 
\begin{gather}\label{eq:26}
\begin{split}
I_{2}= \int_{0}^{\infty} {\frac {\left(b_{N}\right)^2}{x \left( x+b_{N} \right) \left( C_{E} x+b_{N} \right) }} \int_{0}^{x} \frac{t^{k-1} e^{-t}}{(k-1)!} dt\ dx. 
\end{split}
\end{gather}
Changing the order of integration, $I_{2}$ can be rewritten as 
\begin{gather}\label{eq:27}
\begin{split}
I_{2}= \int_{0}^{\infty} {\frac{t^{k-1} e^{-t}}{(k-1)!}\underbrace{ \left( \int_{t}^{\infty} \frac {\left(b_{N}\right)^2}{x \left(x+b_{N} \right) \left( C_{E} x+b_{N} \right) } dx\right) }_\text{$I_{3}(t)$} } dt,
\end{split}
\end{gather}
where $I_{3}(t)$ can be easily evaluated as  
\begin{gather} \label{eq:28}
\begin{split}
I_{3}(t)=
\begin{cases}
-\ln  \left( t \right) +{\frac{C_{E}\ln  \left( t+{\frac { b_{N}}{C_{E}}} \right) -\ln  \left(t+ b_{N} \right)}{C_{E}-1}  }, \ C_{E} \neq 1\\
-\ln  \left( t \right)+ \ln  \left( t+b_{N} \right) -{\frac {b_{N}}{t+b_{N}}}, \ \ \ \ \  C_{E} = 1
\end{cases} .
\end{split}
\end{gather}
Combining (\ref{eq:25}) and (\ref{eq:27}), we have 
\begin{gather} \label{eq:29}
\begin{split}
\overline{C_{s}}{(k,N)} \approx \frac{ 1}{\ln(2)} \int_{0}^{\infty} {\frac{t^{k-1} e^{-t}}{(k-1)!}} I_{3}(t) dt. 
\end{split}
\end{gather}
To evaluate (\ref{eq:29}), let  $V(k;a)= \int_{0}^{\infty}\frac{t^{k-1} e^{-t}}{(k-1)!}\ln(t+a) dt$. Using Eq. (4.337, 5) of \cite{jeffrey2007table}, then we have 
\begin{gather} \label{eq:30}
\begin{split}
V(k;a)= &
\sum_{\mu=0}^{k-1} \frac{ 1 }{ (k-\mu-1)!}  \left( {\left(-1\right)^{k-\mu}  a^{k-\mu-1} e^{a}  E_{i}(-a) }+\right.\\ 
& \left.{ \sum_{v=1}^{k-\mu-1} (v-1)! (-a)^{k-\mu-1-v}}  \right)+ \ln(a),  
\end{split}
\end{gather}
where  $E_{i}(x)=-\int_{-x}^{\infty} \frac{e^{-y}}{y} dy$ is the exponential integral function. From Eq. (4.352, 1)  and  Eq. (3.383, 10)  of  \cite{jeffrey2007table}, we have 
\begin{eqnarray}\label{eq:31}
\int_{0}^{\infty}\frac{t^{k-1} e^{-t}}{(k-1)!} \ln  \left( t \right) dt= \psi(k)
\end{eqnarray}
\begin{eqnarray}\label{eq:32}
\int_{0}^{\infty} \frac{b_{N}}{t+b_{N}}  \frac{t^{k-1} e^{-t}}{(k-1)!} dt= {\left(b_{N}\right)}^{k} {{ e}^{b_{N}}}\Gamma  \left( -k+1,b_{N} \right), 
\end{eqnarray}
respectively. Making use of (\ref{eq:30})-(\ref{eq:32}) in (\ref{eq:29}), we finally obtain $\overline{C_{s}}{(k,N)}$ in (\ref{eq:23}). Using (\ref{eq:23}), we obtain a scaling law for the ESC  in the Corollary below. \\

\noindent \textit{Corollary 1}: For large $N$ relative to a fixed $k$, $\overline{C_{s}}{(k,N)}$ scales as
\begin{equation} \label{eq:33}
\begin{aligned}
\overline{C_{s}}{(k,N)} \sim O\left( \log(N) \right).  
\end{aligned}
\end{equation}
Furthermore, 
\begin{equation} \label{eq:34}
\begin{aligned}
\overline{C_{s}}{(1,N)}-\overline{C_{s}}{(k,N)} \to \frac{H_ {(k-1)}} {\ln(2)} \ \text{bits/s/Hz},  
\end{aligned}
\end{equation}
 where $ H_{(k-1)}=E_{0}+ \psi(k)$ is the harmonic number and $E_{0}=-\psi(1)=0.5772156649$  is the Euler constant. 

\noindent \textit{Proof:} Using the asymptotic behavior of $\Gamma(s,x)\approx x^{s-1} e^{-x}$ for large $x$,  one can show that 
\begin{eqnarray}\label{eq:35}
{\left(b_{N}\right)}^{k} {{ e}^{b_{N}}}\Gamma  \left( -k+1,b_{N} \right) \approx 1, 
\end{eqnarray}
 for large $N$ relative to a fixed $k$. Applying Jensen's inequality:
\begin{gather}\label{eq:36}
\begin{split}
V(k;b_{N})&= \int_{0}^{\infty}\frac{t^{k-1} e^{-t}}{(k-1)!}\ln(t+b_{N}) dt \\
&\leq  \ln\left( \int_{0}^{\infty}\frac{t^{k} e^{-t}}{(k-1)!}  dt+b_{N}\right)\\
&=\ln\left(k+b_{N}\right)\approx \ln\left(b_{N}\right)
\end{split}
\end{gather} 
for large $N$ relative to a fixed $k$. From (\ref{eq:30}) and (\ref{eq:36}), we have $V(k;b_{N}) \approx \ln (b_{N})$ as $ N \to \infty$. Using $V(k;b_{N}) \approx \ln (b_{N})$ and (\ref{eq:35}) with $ b_{N}= C_{M} (N-1)$, (\ref{eq:23}) can be rewritten as 
\begin{gather} \label{eq:37}
\begin{split}
\overline{C_{s}}{(k,N)} \approx
\begin{cases}
\frac{-\psi(k) + \ln \left[C_M(N-1) \right] -{\frac {C_{E}\ln  \left( C_{E} \right) }{C_{E}-1}}
 }{\ln(2)}, \ \    C_{E} \neq 1\\
\frac{-\psi(k) + \ln \left[C_M(N-1) \right] - 1}{\ln(2)}, \ \ \ \  \ \  \ \ \ \  \ \ C_{E} = 1
\end{cases}.  
\end{split}
\end{gather}
From (\ref{eq:37}), we see that $\overline{C_{s}}{(k,N)}  \sim O\left( \log(N) \right)$. Furthermore, $\overline{C_{s}}{(1,N)}-\overline{C_{s}}{(k,N)} \to \frac{ \psi(k)-\psi(1)}{\ln(2)} = \frac{H_ {(k-1)}} {\ln(2)}.$ 

\section{Numerical Results}
Fig. 2, plots the SOP of the $k$-th best user versus the number of users, $N$, for $k=1, 2$, $R_{s}=1, 4$ bit/s/Hz and $L=2$. Some interesting observations can be made: First, we observe that the exact SOP is in good agreement with the simulation results, and the accuracy of the asymptotic SOP in (\ref{eq:12}) increases as $N$ increases. Second, we see that the asymptotic SOP expression is less accurate for small to moderate values of $N$, as $k$ or $R_{s}$ increase. This is due to the fact that the asymptotic analysis holds with a high accuracy for large $N$ compared to fixed $k$ and $R_{s}$. Therefore, if the value of $k$ or $R_{s}$ is close enough to $N$ then the accuracy of  the asymptotic analysis decreases. 

\begin{figure}[h]\label{fig:2}
\begin{center}
\includegraphics[width=1\columnwidth]{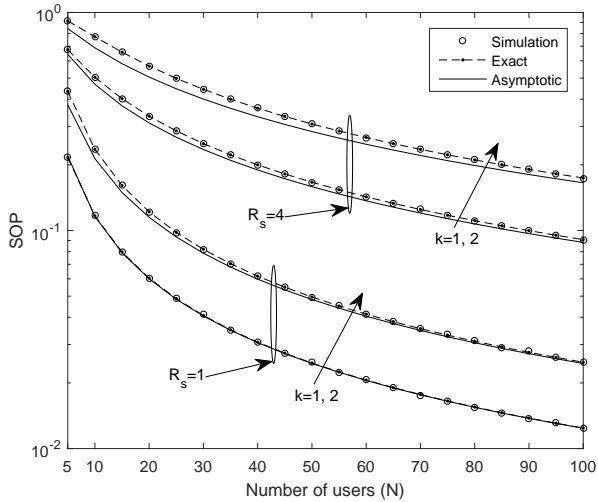}
\caption{SOP of the $k$-th best user vs number of  users for $k=1,2$,  $R_{s}=1, 4$ bit/s/Hz, 
$P/P_{I} = 2$, $\beta_{M} = 2$, $\lambda_{M}= 1/2$, $\beta_{E} = 5$, $\lambda_{E}= 4$ and $L=2$. }
\end{center}
\end{figure} 

\begin{figure}\label{fig:3}
\begin{center}
\includegraphics[width=1\columnwidth]{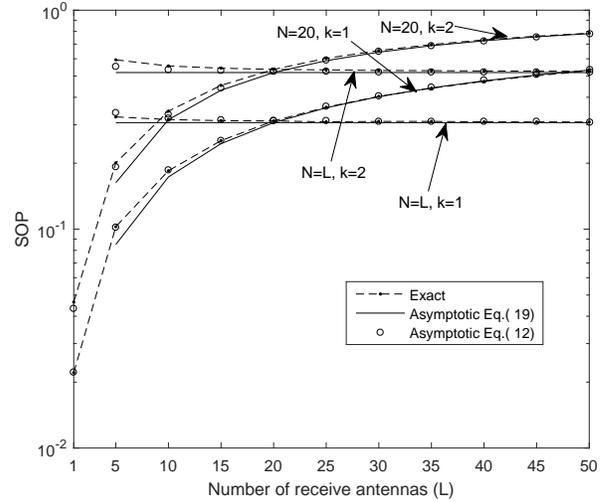}
\caption{SOP of $k$-th best user vs number of receive antennas for $N=20$ and $N=L$, $k=1,2$,  $R_{s}=1/2$ bit/s/Hz, $P/P_{I} = 2$, $\beta_{M} = 2$, $\lambda_{M}= 1/2$, $\beta_{E} = 5$ and $\lambda_{E}= 4$.}
\end{center}
\end{figure} 

\begin{figure}\label{fig:4}
\begin{center}
\includegraphics[width=1\columnwidth]{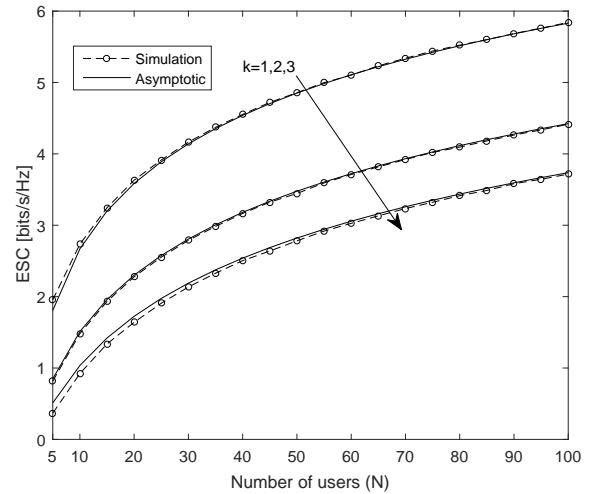}
\caption{ESC of the $k$-th best user vs number of users $N$ for $k=1,2,3$,  
	$P/P_{I} = 4$, $\beta_{M} = 2$, $\lambda_{M}= 4$, $\beta_{E} = 3$ and $\lambda_{E}= 3$ and $L=1$.}
\end{center}
\end{figure} 

In Fig. 3, we plot the SOP of the $k$-th best user as a function of the number of receive antennas, $L$, for $k=1, 2$,  $R_{s}=1/2$ bit/s/Hz and for different values of $N$. We verify the accuracy of the asymptotic SOP expressions derived in (\ref{eq:12}) and (\ref{eq:19}) by comparing them with the exact SOP result. As expected, we observe that for $N=20$, the SOP increases as $L$ increases. However, for $N=L$, the SOP remains constant as $L$ grows large as we discussed earlier at the end of Section III. 

 In Fig. 4, we plot the ESC of the $k$-th best user versus the number of users, $N$, for $k=1, 2, 3$. We validate the accuracy of the asymptotic ESC using Monte Carlo simulations. We observe that the asymptotic ESC is accurate for small to moderate values of $N$. We also observe that asymptotic ESC is less accurate as $k$ approaches $N$. 


\section{Conclusion}
We analyzed the secrecy performance of the $k$-th best user for an interference-limited multiuser network consisting of $N$ legitimate users. We derived closed form exact and asymptotic expressions for the SOP of the $k$-th best user assuming an arbitrary $N$ and large $N$ relative to a fixed $k$, respectively. Furthermore, we derived an asymptotic closed form expression for the ESC of $k$-th best user and showed that the ESC scales like $O\left(\log(N)\right)$ when $N$ grows large relative to a fixed $k$. We also showed that the loss in the ESC between the best user and the $k$-th best user selection converges to a fixed value and it can be quantified by the harmonic number $H_{(k-1)}$. The accuracy of the derived exact and asymptotic expressions were verified, for different system parameters, through Monte Carlo simulations.  


\bibliographystyle{IEEEtran}
\bibliography{yvlc}
\end{document}